\documentclass{sig-alternate}

\usepackage{multirow}
\usepackage{graphicx}
\usepackage{caption}
\usepackage{subcaption}
\usepackage{amsmath}
\usepackage{algorithm2e}
\usepackage[titletoc,toc,title]{appendix}

\newcommand{\loss}{L}
\newcommand{\dec}{\operatorname{dec}}
\newcommand{\cod}{\operatorname{cod}}

\newcommand{\sig}{\operatorname{sig}}

\begin{document}

 \title{
 	Extended Recommendation Framework: Generating the Text of a User Review as a Personalized Summary
}

\numberofauthors{3}
\author{
	\alignauthor
	Micka\"el Poussevin\\
	       \affaddr{University P. et M. Curie, Sorbonne-Universit\'es  Lab. d'Informatique de Paris 6, UMR 7606, CNRS}\\
	       \affaddr{4 Place Jussieu, Paris, France}\\
	       \email{mickael.poussevin@lip6.fr}
	\alignauthor
	Vincent Guigue\\
	       \affaddr{University P. et M. Curie, Sorbonne-Universit\'es  Lab. d'Informatique de Paris 6, UMR 7606, CNRS}\\
	       \affaddr{4 Place Jussieu, Paris, France}\\
	       \email{vincent.guigue@lip6.fr}
	\alignauthor
	Patrick Gallinari\\
	      \affaddr{University P. et M. Curie, Sorbonne-Universit\'es  Lab. d'Informatique de Paris 6, UMR 7606, CNRS}\\
	       \affaddr{4 Place Jussieu, Paris, France}\\
	       	       \email{patrick.gallinari@lip6.fr}
}

\maketitle

\begin{abstract}


We propose to augment rating based recommender systems by providing the user with additional information which might help him in his choice or in the understanding of the recommendation. We consider here as a new task, the generation of personalized reviews associated to items. We use an extractive summary formulation for generating these reviews. We also show that the two information sources, ratings and items could be used both for estimating ratings and for generating summaries, leading to improved performance for each system compared to the use of a single source. Besides these two contributions, we show how a personalized polarity classifier can integrate the rating and textual aspects. Overall, the proposed system offers the user three personalized hints for a recommendation: rating, text and polarity. We evaluate these three components on two datasets using appropriate measures for each task.




\end{abstract}

\begin{figure}
\centering
\includegraphics[width=0.5\textwidth]{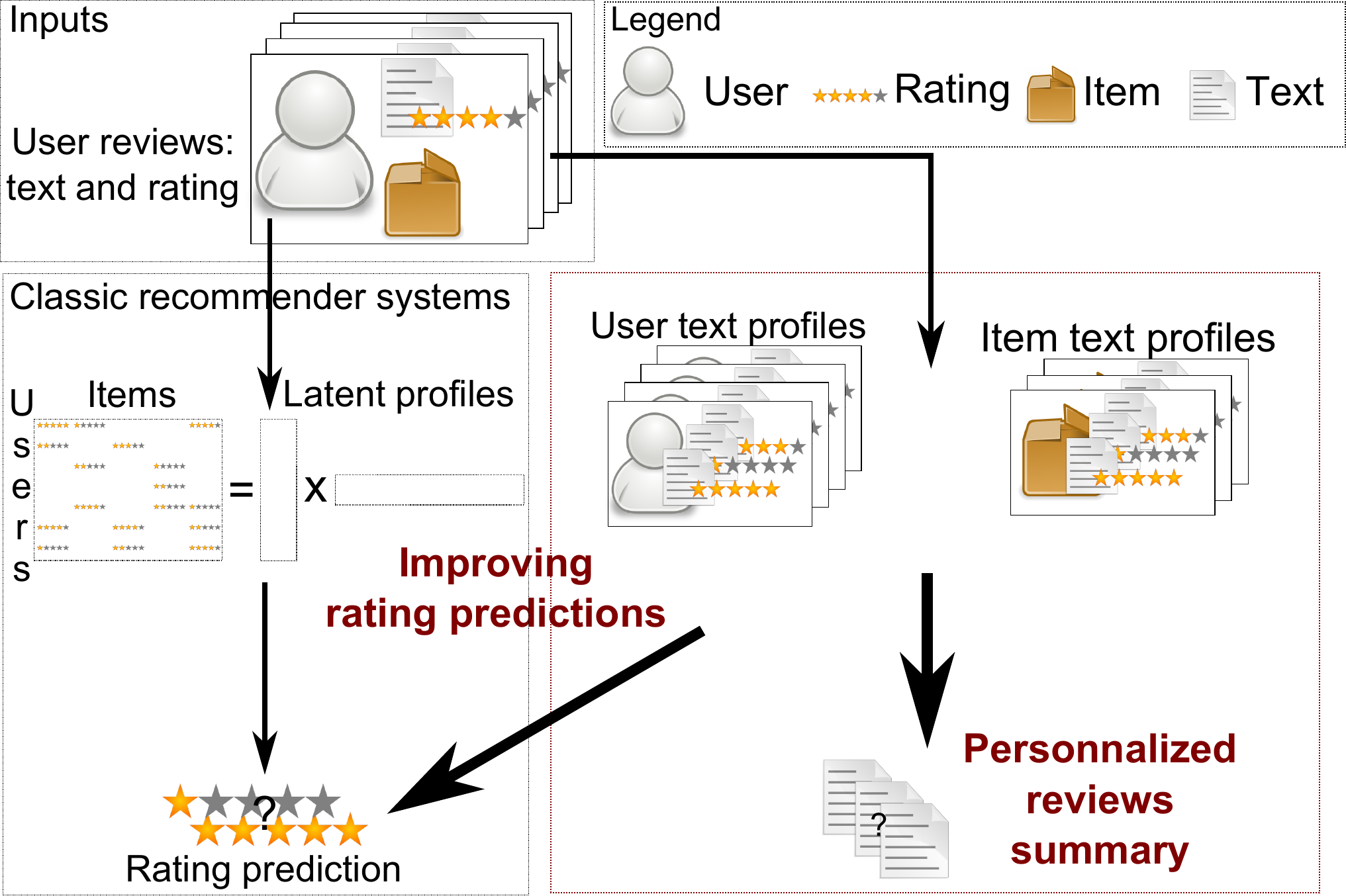}%
\caption{Our contribution is twofold: (1) improving rating predictions using textual information, (2) generating personalized reviews summaries to push recommender systems beyond rating predictions}
\label{fig:model}
\end{figure}

\section{Introduction}
\label{sec:introduction}
The emergence of the participative web has enabled users to easily give their sentiments on many different topics. 
This opinionated data flow thus grows rapidly and offers opportunities for several applications like e-reputation management or recommendation.
Today many e-commerce websites present each item available on their platform with a description of its characteristics, average appreciation, ratings together with individual user reviews explaining their ratings.


Our focus here is on user - item recommendation. This is a multifaceted task where different information sources about users and items could be considered and different recommendation information could be provided to the user.
Despite this diversity, the academic literature on  recommender systems has focused only on a few specific tasks. The most popular one is certainly the prediction of user preferences given their past rating profile. These systems typically rely on collaborative filtering \cite{Koren2009} to predict missing values in a \emph{user/item/rating} matrix. In this perspective of rating prediction, some authors have make use of additional information sources available on typical e-commerce sites. \cite{Ganu2009} proposed to extract topics from  consumer reviews in order to improve ratings predictions. Recently, \cite{McAuley2013a}  proposed to learn a latent space common to both textual reviews and product ratings, they showed that rating prediction  was improved by such hybrid recommender systems. Concerning the information provided to the user, some models exploit review texts for ranking comments that users may like \cite{Agarwal2011} or for answering specific user queries \cite{tan2012each}.

We start here from the perspective of predicting user preference  and argue that the exploitation of the information present in many e-commerce sites, allows us to go beyond simple rating prediction for presenting  users with complementary information that may help him making his choice. We consider as an example the generation of a personalized review accompanying each item recommendation. Such a review is a source of complementary evidence for the user appreciation of a suggestion. Similarly as it is done for the ratings, we exploit past information and user similarity in order to generate these reviews. Since pure text generation is a very challenging task \cite{Amini2007a}, we adopt an extractive summary perspective: the generated text accompanying each rating will be extracted from the reviews of selected users who share similar tastes and appreciations with the target user. Ratings and reviews being correlated, this aspect could also be exploited to improve the predictions. Our rating predictor will make use of user textual profiles extracted from their reviews and summary extraction in turn will use predicted ratings. Thus both types of information, predicted ratings and generated text reviews, are offered to the user and each prediction, rating and generated text, takes into account the two sources of information. Additional information could also be provided to the user. We show here as an example, that predicted ratings and review texts can be used to train a robust sentiment classifier which provides the user with a personalized polarity indication about the item.
The modules of our  system are evaluated on the two main tasks, rating prediction and summary extraction, and on the secondary task of sentiment prediction. For this, experiments are conducted on real datasets collected from \emph{amazon.com} and \emph{ratebeer.com} and models are compared to classical baselines.

The recommender system is compared to a classic collaborative filtering model using the mean squared error metric. We show that using both ratings and user textual profiles allows us to improve the performance of a baseline recommender. 
Gains are motivated from a more precise understanding of the key aspects and opinions included in the item and user textual profiles.
For evaluating summary text generation associated to a couple (user, item), we have at our disposal a gold standard, the very review text written by this user on the item. Note that this is a rare situation in summary evaluation. However contrarily to collaborative filtering, there is no consensual baseline. We then compare our results to a random model and to oracle optimizing the ROUGE-n metric. They respectively provide a lower and an upper bound of the attainable performance. The sentiment classifier is classically evaluated using classification accuracy.

This article is organized as follows. 
The hybrid formulation, the review generator and the sentiment classifier are presented in section~\ref{sec:models}.
Then, section~\ref{sec:experiments} gives an extensive experimental evaluation of the framework.
The overall gains associated to hybrid models are discussed in section~\ref{sec:overall gains}.
A review of related work is provided in section~\ref{sec:related work}.

\section{Models}
\label{sec:models}
In this section, after introducing the notations used throughout the paper, we will describe successively the three modules of our system.
We start by considering the prediction of ratings \cite{McAuley2013a}. Rating predictors answer the following question: \emph{what rating will this user give to this item?} We present a simple and efficient way to introduce text profiles representing the writing style and taste of the user in a hybrid formulation. We then show how to exploit reviews and ratings in a new challenging task: \emph{what text will this user write about this item?} We propose an extractive summary formulation of this task. We then proceed to describe how both ratings and text could be used together in a personalized sentiment classifier.

\subsection{Notations}
\label{sec:models:subsec:notations}

We use $u$ (respectively $i$) to refer to everything relative to a user (respectively to an item) and the rating given by user $u$ to the item $i$ is denoted $r_{u i}$.
$U$ and $I$ refer to anything relative to all users and all items, such as the rating matrix $R_{UI}$.
Similarly, lower case letters are used for scalars or vectors and upper case letters for matrices.
$d_{ui}$ is the actual review text  written by user $u$ for item $i$. 
It is composed of $\kappa_{ui}$ sentences: $d_{ui} = \{s_{uik}, 1 \leq k \leq \kappa_{ui}\}$.
In this work, we consider documents as bags of sentences. To simplify notations, $s_{uik}$ is replaced by $s_{ui}$ when there is no ambiguity.
Thus, user appreciations are quadruplets $(u, i, r_{u i}, d_{u i})$. 
Recommender systems use past information to compute a rating prediction $\hat{r}_{ui}$, the corresponding prediction function is denoted $f(u, i)$.

For the experiments, ratings and text reviews are split into training, validation and test sets 
respectively denoted $S_{train}$, $S_{val}$ and $S_{test}$ and containing $m_{train}$, $m_{val}$ and $m_{test}$ user appreciations (text and rating).
We denote $S_{train}^{(u)}$, defined in equation \eqref{eq:def strainu}, the subset of all reviews $S_{train}$ that were written by user $u$ and $m_{train}$ the number of such reviews.
\begin{equation}
	\label{eq:def strainu}
	S_{train}^{(u)} = \{(u', i', r_{u'i'}, d_{u'i'}) \in S_{train}, u' = u\}
\end{equation}
Similarly, $S_{train}^{(i)}$ and $m_{train}^{(i)}$ are used for the reviews on item $i$.
In order to simplify notations in the following section, we denote $\sum_{S_{train}}$ the sum where all quadruplet from the training set are considered.

\subsection{Hybrid recommender system with text profiles}
\label{sec:models:subsec:recommender systems with text profiles of users}

Recommender systems classically use rating history to predict the rating $\hat{r}_{ui}$ that user $u$ will give to item $i$.
The hybrid system described here makes use of both \emph{collaborative filtering} through matrix factorization and textual information to produce a rating as described in~\eqref{eq:mixture recommender systems}:

\begin{eqnarray}
	f(u, i)    &=& \mu + \mu_u + \mu_i + \gamma_u.\gamma_i + g (u, i) \label{eq:mixture recommender systems} 
\end{eqnarray}

The first three predictors  in equation \eqref{eq:mixture recommender systems} are biases (overall bias, user bias and item bias).
The fourth predictor is a classical matrix factorization term.
The novelty of our model comes from the fifth term~\eqref{eq:mixture recommender systems} that takes into account text profiles to refine the prediction $f$.
Our aim for the rating prediction is to minimize the following empirical loss function:
\begin{eqnarray}
 \underset{\mu,\mu_{u},\mu_{i},\gamma_{u},\gamma_{i},g }{\operatorname{argmin}} \loss &=& \frac{1}{m_{train}}\sum_{ S_{train}} \left(r_{ui} - f(u, i)\right)^{2} \label{eq:loss recommender system}
\end{eqnarray}
To simplify the learning procedure, we first optimize the parameters of the different components independently as described in the following subsections. 
Then we fine tune the combination of these components by learning weighting coefficients so as to maximize the performance criterion~\eqref{eq:mixture recommender systems} on the validation set. This procedure is described in section~\ref{sec:mixture}.

\subsubsection{Matrix factorization}
\label{sec:factomat}

We first compute the different bias from eq.~\eqref{eq:mixture recommender systems} as the averaged ratings over their respective domains (overall, user  and item):

\begin{eqnarray}
	\mu^* = \frac{1}{m_{train}}\sum_{ S_{train}} r_{ui} \\
	\mu_u^* = \frac{1}{m_{train}^{(u)}} \sum_{ S_{train}^{(u)}} r_{ui}, &
	\displaystyle \mu_i^* = \frac{1}{m_{train}^{(i)}} \sum_{ S_{train}^{(i)}} r_{ui} 
\end{eqnarray}

For the matrix factorization term,
we approximate the rating matrix $R_{UI}$ using two latent factors: $R_{UI} \approx \Gamma_U\Gamma_I^t$.
Both $\Gamma_U$ and $\Gamma_I$ are two matrices representing collections of latent profiles, with one profile per row.
The former, $\Gamma_U$, contains the latent representation corresponding to the taste of each user.
The latter, $\Gamma_I$, contains the latent profile of each item.
We denote $\gamma_u$ (resp. $\gamma_i$) the row of $\Gamma_U$ (resp. $\Gamma_I$) corresponding to the latent profile of user $u$ (resp. item $i$).


The profiles are learned by minimizing, on the training set, the mean squared error between known ratings in matrix $R_{UI}$ and the approximation provided by the factorization $\Gamma_U \Gamma_I^T$.
This minimization problem described in equation \eqref{eq:mf problem loss}, with an additional L2 constraint \eqref{eq:mf problem regularization} on the factors is solved here using non-negative matrix factorization.

\begin{eqnarray}
	\Gamma_U^*, \Gamma_I^* = \underset{\Gamma_U, \Gamma_I}{\operatorname{argmin}} && \| M_{train} \odot(R_{UI} - \Gamma_U\Gamma_I)\|_F^2 \label{eq:mf problem loss}\\
	&&+ \lambda_U\|\Gamma_U\|_F^2 + \lambda_I\|\Gamma_I\|_F^2 \label{eq:mf problem regularization}
\end{eqnarray}

In this equation $M_{train}$ is a mask that has the same dimensions as the rating matrix $R_{UI}$,  an entry  is 1 only if the corresponding review is in the training set and zero otherwise and $\odot$ is the element-wise product on matrices.

\subsubsection{Text profiles exploitation}
\label{sec:models:subsubsec:text profiles exploitation}

Let us denote $\pi_u$ the profile of user $u$ and $\sigma_{t}(\pi_{u'}, \pi_u)$ a similarity operator between user profiles.
The last component of the predictor $f$ in \eqref{eq:mixture recommender systems} is a weighted average of user ratings for item $i$, where weight $\sigma_t(\pi_{u'}, \pi_u)$ is the similarity between the text profiles $\pi_{u'}$ and $\pi_u$ of users $u'$ and $u$, the latter being the target user.
This term takes into account the fact that two users with similar styles or using similar expressions in their appreciation of an item, should share close ratings on this item.
The prediction term for the user/item couple $(u, i)$ is then expressed as \eqref{eq:text profiles prediction}:

\begin{equation}
	\label{eq:text profiles prediction}
	g(u, i) = \frac{1}{m_{train}^{(i)}} \sum_{S_{train}^{(i)}} r_{u'i} \sigma_t(\pi_u', \pi_u)
\end{equation}

Two different representations for the text profiles $\pi_u$ of the users are investigated in this article: one is based on a latent representation of the texts obtained by a neural network autoencoder, the other relies on a robust bag of words coding. Each one is associated to a dedicated metric $\sigma_{t}$.

This leads to two formulations of $g$, and thus, to two rating prediction models. 
We denote the former $f_{A}$ (autoencoder) and the latter $f_T$ (bag of words). Details are provided below.

\paragraph{Bag of words}

A preprocessing step removes all words appearing in less than 10 documents.
Then, the $100\,000$ most frequent words are kept.
Although the number of features is large, the representation is sparse and scales well. 
$\pi_u$ is simply the binary bag of words of all texts of user $u$. 
In this high dimensional space, the proximity in style between two users is well described by a cosine function, a high value indicates similar usage of words: 
\begin{equation}
	\label{eq:text profiles cosine similarity}
	\sigma_t(\pi_{u'}, \pi_u) = \pi_{u'}  \pi_u / (\|\pi_{u'}\|\|\pi_u\|)
\end{equation}

\paragraph{Autoencoder} 
%
The neural network autoencoder has two components: a coding operator and a decoding operator denoted respectively $cod$ and $dec$. The two vectorial operators are learned so as to enable the reconstruction of the original text after a projection in the latent space.
Namely, given a sentence $s_{uik}$ represented as a binary bag of words vector, we obtain a latent profile $\pi_{s_{uik}} = cod(s_{uik})$ and then, we reconstruct an approximation of the sentence using $\hat s_{uik} = dec(\pi_{s_{uik}})$.

The autoencoder is optimized so as to minimize the reconstruction error over the training set:
%
%
\begin{equation}
	\label{eq:loss text representation}
\begin{array}{ll}
cod^{*},dec^{*} = \\ \quad \displaystyle \underset{cod,dec}{\operatorname{argmin}}  \sum_{ S_{train}} \frac{1}{\kappa_{ui}} \sum_{k = 1}^{\kappa_{ui}} \| s_{uik} - dec(cod(s_{uik})) \|^2 
\end{array}
\end{equation}

We use the settings proposed in \cite{glorot2011domain}: our dictionary is obtained after stopwords removal and selecting the most frequent 5000 words. we did not use a larger dictionary such as the one used for the bag of word representation since it does not lead to improved performance and  simply increases the computational load.
All sentences are represented as binary bag of words using this dictionary.
The coding dimension has been set to 1000 after a few evaluation trials. Note that the precise value of this latent space is not important and the performance is similar on a large range of dimension values.
Both $cod$ and $dec$ use sigmoid units:
\begin{equation}
	\begin{array}{ll}
		\displaystyle \cod(s_{uik}) = \pi_{uik} = \sig(Ws_{uik} + b) \\
		\displaystyle \dec(\pi_{uik}) = \sig(W^{T} \pi_{uik}+b') \\
		\displaystyle \sig(t) = \frac{1}{1 + \exp(-t)}
	\end{array}
\end{equation}

Here, $\pi_{uik}$ is a vector, $W$ is a 5000x1000 weight matrix and $\sig()$ is a pointwise sigmoid operator operating on the vector $Ws_{uik} + b$.

As motivated in \cite{McAuley2013a,Ganu2009}, such a latent representation helps exploiting term co-occurrences and thus introduces some semantic.
It provides a robust text representation. 
The hidden activity of this neural network produces a continuous representation for each sentence accounting for the presence or absence of groups of words.

$\pi_u$ is obtained by coding the vector corresponding to all text written by the user $u$ in the past.
It lies in a latent word space where a low Euclidean distance between users means a similar usage of words.
Thus, for the similarity $\sigma_t$, we use an inverse Euclidean distance in the latent space: 
\begin{equation}
	\sigma_t(\pi_{u'}, \pi_u) = 1/(\alpha + \|\pi_{u'}- \pi_u\|)
\end{equation}


\subsubsection{Global training criterion for ratings prediction}
\label{sec:mixture}

 In order to connect all the elementary components described above with respect to our recommendation task, we introduce  weighting parameters $\beta$ in~\eqref{eq:mixture recommender systems}.
Thus, the initial optimization problem \eqref{eq:loss recommender system} becomes:
 \begin{eqnarray}
\begin{array}{ll}
\beta^{*} = \underset{\beta }{\operatorname{argmin}} \frac{1}{m_{train}}\sum_{ S_{train}} \\
 \Big(r_{ui} - \big(\beta_{1} \mu^{*} + \beta_{2}\mu_u^{*} + \beta_{3}\mu_i^{*} + \beta_{4}\gamma_u^{*}.\gamma_i^{*} + \beta_{5}g (u, i)\big)\Big)^{2} 
\end{array}
  \label{eq:globlin recommender system}
\end{eqnarray}

 The linear combination is optimized using a validation set: this step guaranties that all components are combined in an optimal manner.


\subsection{Text generation model}
\label{sec:models:subsec:text generation model}

The goal here is to generate a review text for each ($u$,$i$) recommendation. During the recommendation process, this text is an additional information for users to consider. It should catch their interest and in principle be close to the one that user $u$ could have written himself on item $i$. Each text is generated as an extractive summary, where the extracted sentences $s_{u'i}$ come from the reviews written by other users ($u' \neq u$) about item $i$. Sentence selection is performed according to a criterion which combines a similarity between the sentence and the textual user profile and a similarity between the actual rating $r_{u'i}$ and the prediction made for ($u$,$i$),  $\hat r_{ui}$ computed as described in section 2.2. The former measure could take into account several dimensions like vocabulary, sentiment expression and even style, here it is mainly the vocabulary which is exploited. The latter measures the proximity between user tastes.  
For the text measure, we make use of  the $\sigma_t$ similarity  introduced in section~\ref{sec:models:subsec:recommender systems with text profiles of users}. 
As before, we will consider two representations for texts (latent coding and raw bag of words).
For the ratings similarity, we use $\sigma_r(r_{u'i} , r_{ui}) = 1/(1+|r_{u'i} - r_{ui}|)$.

Suppose one wants to select a single sentence for the extracted summary. The sentence selection criterion will then be a simple average of the two similarities:

\begin{equation}
	\label{eq:scoring model}
	h(s_{u' i}, r_{u' i}, u', u, i) = \frac{\sigma_t(s_{u' i}, \pi_u) + \sigma_r(r_{u' i}, \hat{r}_{ui})}{2} 
\end{equation}

Note that this function may score any piece of text. 
In the following, we then consider three possibilities for generating text reviews:
The first one simply consists in selecting the best sentence $s_{u'i}$ among all the training sentences for item $i$ with respect to $h$. We call it 1S for single sentence. 
The second one selects a whole review $d_{u'i}$ among all the reviews for $i$. The document is here considered as one long sentence. 
This is denoted CT for complete text. 
The third one is a greedy procedure that selects multiple sentences, it is denoted XS. 
It is initialized with 1S, and then sentences are selected under two criteria: relevance with respect to $h$ and diversity with respect to the sentences already selected. 
Selection is stopped when the length of the text is greater than the average length of the texts of the target user.
Algorithm~\ref{algo:selsent} sums up the XS procedure for generating the text $\hat{d}_{ui}$ for the couple user $u$, item $i$. 

\begin{algorithm}
	\SetAlgoLined
	\KwData{$u$, $i$, $S =\{(s_{u'i}, r_{u'i}\ u'\}$}
	\KwResult{$\hat{d}_{ui}$}
	$s^*_{u'i} \gets \underset{s_{u'i} \in S}{\operatorname{argmax}} \big(h(s_{u'i}, r_{u'i}, u', u, i)\big)$\;
	$\hat{d}_{ui} \gets s^*_{u'i}$\;
	Remove $s^*_{u'i}$ from $S$\;
	\While{$\operatorname{length}{\hat{d}_{ui}} < \operatorname{averagelength}(u)$}{
		$s^*_{u'i} \gets \underset{s_{u'i} \in S}{\operatorname{argmax}} \big( h(s_{u'i}, r_{u'i}, u', u, i) - \cos(s_{u'i}, \hat{d}_{ui})\big)$ \;
		$\hat{d}_{ui} \gets s^*_{u'i}$\;
		Remove $s^*_{u'i}$ from $S$\;
	}
	\caption{\label{algo:selsent}XS greedy procedure: selection of successive sentences to maximize both relevance and diversity. $\hat{d}_{ui}$ is the text that is generated, sentence after sentence.}
\end{algorithm}

\subsection{Sentiment prediction model}
\label{sec:models:subsec:sentiment prediction model}

Besides ratings and text reviews, other additional information might help the user for his choice. We show here as an example  how polarity information about an item could be generated by exploiting both the user predicted ratings and his textual profile. It will be shown that exploiting both information sources improves the sentiment prediction performance compared with a usual text based sentiment classifier.
 

Polarity classification \cite{Pang2002} is the task of predicting whether a text $d_{ui}$ (here of a review) is positive or negative.
We use as ground truth the ratings $r_{ui}$ and follow a standard thresholding procedure \cite{Pang2008}: reviews rated $1$ or $2$ are considered as negative, while items rated $4$ or $5$ are positive. 
All texts that are rated $3$ are ignored as it is unclear whether that are positive or negative: it strongly depends on the rating habits of the user. Note that more sophisticated thresholdings could be considered as well, but we used this one here since it is very common.


As with the rating predictor and the text generator, this sentiment classifier will be evaluated on our test data.
For this, one considers two baselines. A first one only uses the rating prediction of our recommender system $f(u,i)$ as a label prediction, this value is then thresholded as indicated above. A second one is a classical text sentiment classifier. Denoting by $d_{ui}$ the binary bag of word representation of a document and $c_{ui}$ the binary label associated to the rating $r_{ui}$, one uses a linear SVM $s(d_{ui}) = d_{ui} .w$. Note that this is usually a strong baseline for the polarity classification task.
Our final classifier will combine $f(u,i)$ and $s(d_{ui})$ in order to solve the following optimization problem:

\begin{equation}
	\label{eq:loss sentiment prediction}
\begin{array}{ll}
\displaystyle	w^* = \\ \ \underset{w}{\operatorname{argmin}} \sum_{ S_{train}, r_{ui} \neq 3} \Big(1- \big(d_{ui}.w + f(u,i)\big) c_{ui} \Big)_{+} + \lambda \|w\|^{2} \\
	\text{with } (x)_{+} = \left\{ 	
\begin{array}{ll}
 x &\text{ if } x>0\\
 0 &\text{ otherwise}
\end{array}
		\right.
\end{array}
\end{equation}

Note that here only $w$ is learned and that for $f$ one simply considers the recommender system trained earlier.
In the experimental section, we will also compare the results obtained with the two versions of our rating predictor: $f_{T}$ and $f_{A}$ (cf section~\ref{sec:models:subsubsec:text profiles exploitation}).


\section{Experiments}
\label{sec:experiments}

All three modules, ratings, text, sentiments, are evaluated independently since there is no global evaluation framework. These individual performances should however provide together a quantitative appreciation of the whole system.

We use two real world datasets of user reviews, collected from \textit{amazon.com} \cite{jindal2008opinion} and \textit{ratebeer.com} \cite{McAuley2013a}. Their characteristics are presented in table~\ref{tab:datasets information}.

Below, one presents first how datasets are preprocessed in \ref{sec:experiments:subsec:data preprocessing}.
The benefits of incorporating the text in the ratings prediction for the recommender system are then discussed  in section~\ref{sec:experiments:subsec:recommender system evaluation}.
The quality of the generated reviews is evaluated and analyzed in section~\ref{sec:experiments:subsec:text generation evaluation}
Finally, the performance of the sentiment classifier combining text and ratings is described in~\ref{sec:experiments:subsec:sentiment classification evaluation}.

\begin{table*}[tb!]
	\centering
	\begin{tabular}{|c|c|c|c|c|c|c|}
		\hline
		 Source &           Subset names      & \#Users & \#Items & \multicolumn{3}{c|}{\#Reviews} \\ &&&&\#Training & \#Validation & \#Test \\
		\hline
		\multirow{3}{*}{\rotatebox{45}{Ratebeer}} & RB\_U50\_I200      & 52             & 200      & 7200           & 900          & 906    \\
		&RB\_U500\_I2k      & 520            & 2000     & 388200         & 48525        & 48533  \\
		&RB\_U5k\_I20k      & 5200           & 20000    & 1887608        & 235951       & 235960 \\
		\hline
		\multirow{4}{*}{\rotatebox{45}{Amazon}}&A\_U200\_I120      & 213            & 122      & 984            & 123          & 130    \\
		&A\_U2k\_I1k        & 2135           & 1225     & 31528          & 3941         & 3946   \\
		&A\_U20k\_I12k      & 21353          & 12253    & 334256         & 41782        & 41791  \\
		&A\_U210k\_I120k    & 213536         & 122538   & 1580576        & 197572       & 197574 \\
		\hline
	\end{tabular}
	\caption{\label{tab:datasets information}Users, items \& reviews counts for every datasets.}
	\centering
	\begin{tabular}{|l||c|c|c|c||c|c|}
		\hline
		Subsets           & $\mu$      & $\mu_u$       & $\mu_i$       & $\gamma_u . \gamma_i$ & $f_A$                & $f_T$     \\ 
		\hline
		RB\_U50\_I200     & 0.7476   & 0.7291      & 0.3096      & 0.2832              & \textbf{0.2772}    & \textbf{0.2773}    \\
		RB\_U500\_I2k     & 0.6536   & 0.6074      & 0.3359      & 0.3168              & \textbf{0.3051}    & \textbf{0.3051}    \\
		RB\_U5k\_I20k     & 0.7559   & 0.6640      & 0.3912      & 0.3555              & \textbf{0.3451}    & \textbf{0.3451}    \\
		\hline
		A\_U200\_I120     & 1.5348   & 2.0523      & 1.6563      & 1.7081              & \textbf{1.4665}    & 1.4745    \\
		A\_U2k\_I1k       & 1.5316   & 1.4391      & 1.3116      & 1.0927              & \textbf{1.0483}    & \textbf{1.0485}    \\
		A\_U20k\_I12k     & 1.4711   & 1.4241      & 1.2849      & 1.0797              & \textbf{1.0426}    & \textbf{1.0426}    \\
		A\_U210k\_I120k   & 1.5072   & 2.1154      & 1.5318      & 1.2915              & \textbf{1.1671}    & \textbf{1.1678}    \\
		\hline
	\end{tabular}
	\caption{
		\label{tab:mse all}
		Test performance (mean squared error) for recommendation.
		$\mu$, $\mu_u$, $\mu_i$ are the overall bias, user bias and item bias baselines.
		$\gamma_u . \gamma_i$ is the plain matrix factorization baseline.
		$f_A$, $f_T$ are our hybrid recommender systems relying respectively on latent and raw text representations. The different datasets are described in table~\ref{tab:datasets information}.
		}
\end{table*}

\subsection{Data preprocessing}
\label{sec:experiments:subsec:data preprocessing}
Reviews from different websites have different formats (rating scales, multiple ratings, \dots), they are then preprocessed to a unified format. 
Ratings are scaled to a 1 to 5 integer range.
For titled reviews, the title is considered as the first sentence of the text of the review.
Each dataset is randomly split into three parts: training, validation and test containing respectively 80\%, 10\% and 10\% of the reviews.

As described in \ref{sec:models:subsec:recommender systems with text profiles of users}, two representations of the text are considered each with a different dictionary:

\begin{itemize}
	\item for the autoencoder, we have selected the 5000 most frequent words, with a stopwords removal step; The autoencoder input vector is then a binary vector of dimension 5000.
	\item for the raw representation, we have selected the 100000 most frequent words appearing in more than 10 documents (including stopwords) and used a binary vector representation.
\end{itemize}

For the experiments, we consider several subsets of the databases with different numbers of users and items. 
Each dataset is built by extracting, for a given number of users and items, the most active users and the most commented items. 
Dataset characteristics are given in table \ref{tab:datasets information}.

\begin{table*}[htpb!]
	\centering
	\begin{tabular}{|l||c||c|c|c|c||c|c|}
		\hline
		Subsets           & LL    & $\mu_i$ & $\gamma_u . \gamma_i$ & $f_A$ & $f_T$ & LL + $f_A$  & LL + $f_T$  \\
		\hline
		RB\_U50\_I200     & 5.35  &  5.12   &  6.01                 &  5.57 & 5.57  & \textbf{3.79}        & \textbf{3.79}        \\
		RB\_U500\_I2k     & 7.18  & 10.67   & 9.73                  &  8.55 & 8.55  & \textbf{6.52}        & 6.92        \\
		RB\_U5k\_I20k     & 8.44  & 11.80   & 10.04                 &  9.17 & 9.17  & \textbf{8.33}        & \textbf{8.35}        \\
		\hline
		A\_U200\_I120     & \textbf{10.00} & 15.83   & 22.50                 & 20.00 & 20.83 & \textbf{10.00}       & \textbf{10.00}       \\
		A\_U2k\_I1k       &  7.89 & 15.25   & 12.85                 & 12.62 & 12.62 &  \textbf{7.54}       &  \textbf{7.54}       \\
		A\_U20k\_I12k     &  \textbf{6.34} & 13.99   & 12.79                 & 12.38 & 12.37 &  \textbf{6.29}       &  \textbf{6.29}       \\
		A\_U210k\_I120k   &  \textbf{6.25} & 14.04   & 14.40                 & 13.32 & 13.31 &  \textbf{6.22}       &  \textbf{6.22}       \\
		\hline
	\end{tabular}
	\caption{\label{tab:ce all}Test performance (classification error) as polarity classifiers.
	LL stands for LibLinear (SVM), $\mu_i$, $\gamma_u . \gamma_i$, $f_A$, $f_T$ are the recommender systems as in table \ref{tab:mse all}. 
	LL + $f_A$  and LL + $f_T$ are two hybrid opinion classification models combining the SVM classifier and $f_A$ and $f_T$ recommender systems.}
\end{table*}

\subsection{Recommender system evaluation}
\label{sec:experiments:subsec:recommender system evaluation}

Let us first consider the evaluation of the rating prediction.
The metric used here is the mean squared error (MSE) between rating predictions $\hat{r}_{ui}$ and actual ratings $r_{ui}$. 
The lower the MSE is, the better the model is able to estimate the correspondence between user tastes and items.
Results are presented in table~\ref{tab:mse all}.

The models are referenced using the notations introduced in section~\ref{sec:models:subsec:recommender systems with text profiles of users}.
The first column corresponds to a trivial system which predicts $\mu$ the overall bias, the second predicts the user bias $\mu_u$. Both give poor performance as expected.

The third column corresponds to the item bias $\mu_i$ baseline. 
It assumes that user taste is not relevant and that each item has its own intrinsic quality. The improvement with respect to $\mu$ and $\mu_u$ is important since MSE is halved.
The fourth column corresponds to a nonnegative matrix factorization baseline, denoted $\gamma_u . \gamma_i$.
It jointly computes latent representations for user tastes and items characteristics. Unsurprisingly, it is our best baseline.

It could be noted that performance tends to degrade when the subset size increases.
This is a side effect associated to the review selection process used for building the different datasets. 
Smaller datasets contain the most active users and the most commented items.
The estimation of their profiles benefits from the high number of reviews per user (and item) in this context.

The last two columns refer to our hybrid recommender systems, using the two text representations introduced in section~\ref{sec:models:subsec:recommender systems with text profiles of users}.
Both $f_{A}$ (autoencoder) and $f_{T}$ (raw text) perform better than a baseline collaborative filtering system and both have similar approximation errors.
The main difference between the systems comes from the complexity of the approach: during the learning step, $f_T$ is much faster than $f_{A}$ given the fact that no autoencoder optimization is required. On top of that, $f_T$  remains faster in the inference step:
the inherent sparsity of the bag of word representation enables $f_{T}$ to provide faster computations than $f_{A}$. The autoencoder works in a smaller dimensional space but it is not sparse.


\subsection{Text generation evaluation}
\label{sec:experiments:subsec:text generation evaluation}

We move on now to the evaluation of the personalized review text generation module. Since we are using an extractive summary procedure, we make use of a classical loss used for summarization systems. The quality of the prediction is thus evaluated with a recall-oriented ROUGE-n metrics, by comparing the generated text against the actual text of the review produced by the user. As mentioned before, this is a rare case when a generated summary could be compared with the one actually written by a user. As far as we know, generating candidate reviews has never been dealt with in this context and this is a novel task. 
The ROUGE-n metric is the proportion of n-grams of the actual text found in the predicted (candidate) text.
Here, three metrics are used: ROUGE-1, ROUGE-2 and ROUGE-3.
The higher ROUGE-n is, the better the quality of the candidate text is.
A good ROUGE-1 means that topics or vocabulary are correctly caught while ROUGE-2 and ROUGE-3 are more representative of the discussed topics.

\begin{figure*}[tb!]
\centering
\begin{subfigure}[b]{0.45\textwidth}\includegraphics[scale=0.4]{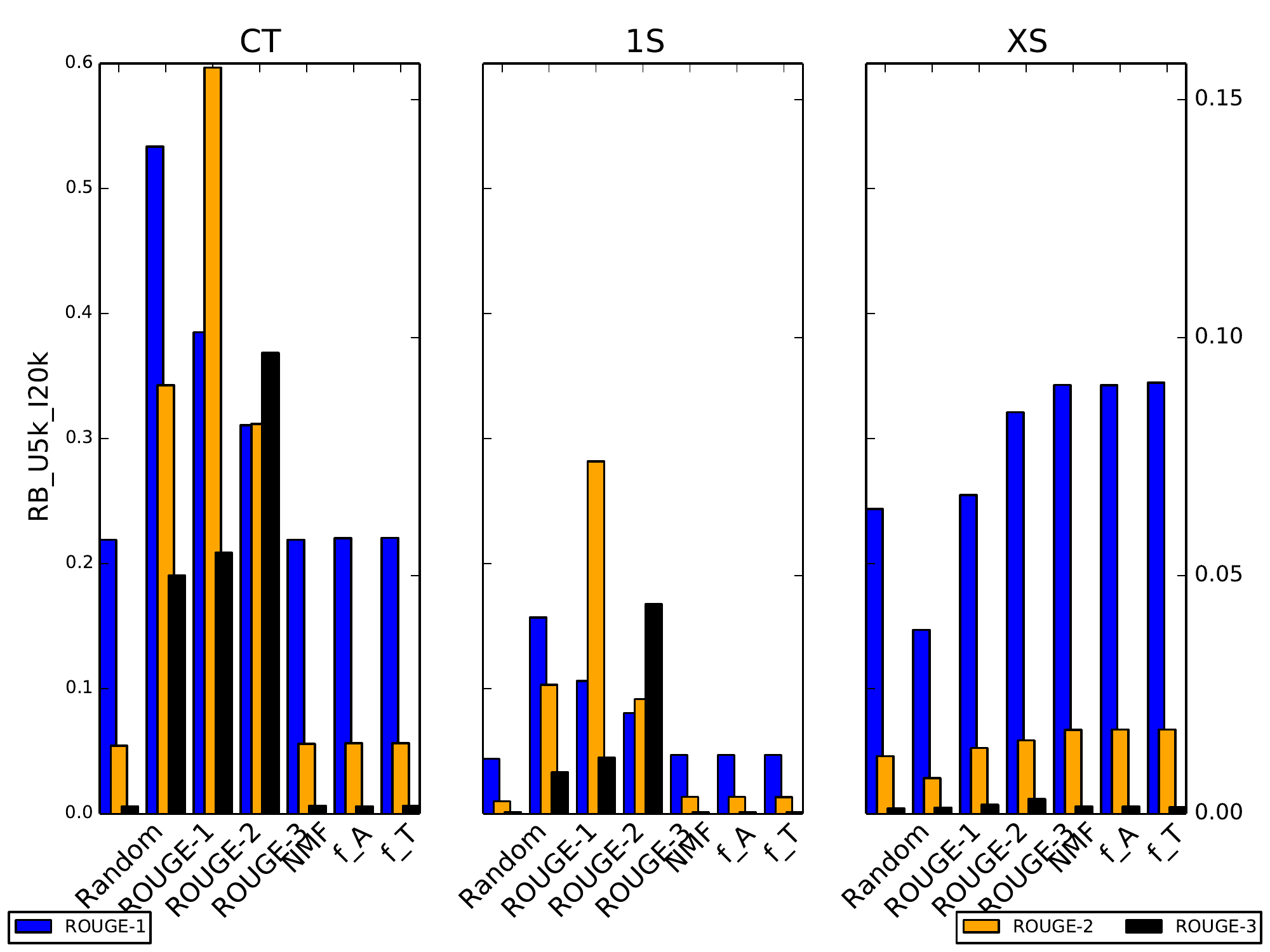}\caption{\label{fig:rouge ratebeer}RateBeer experiments}\end{subfigure} ~
\begin{subfigure}[b]{0.45\textwidth}\includegraphics[scale=0.4]{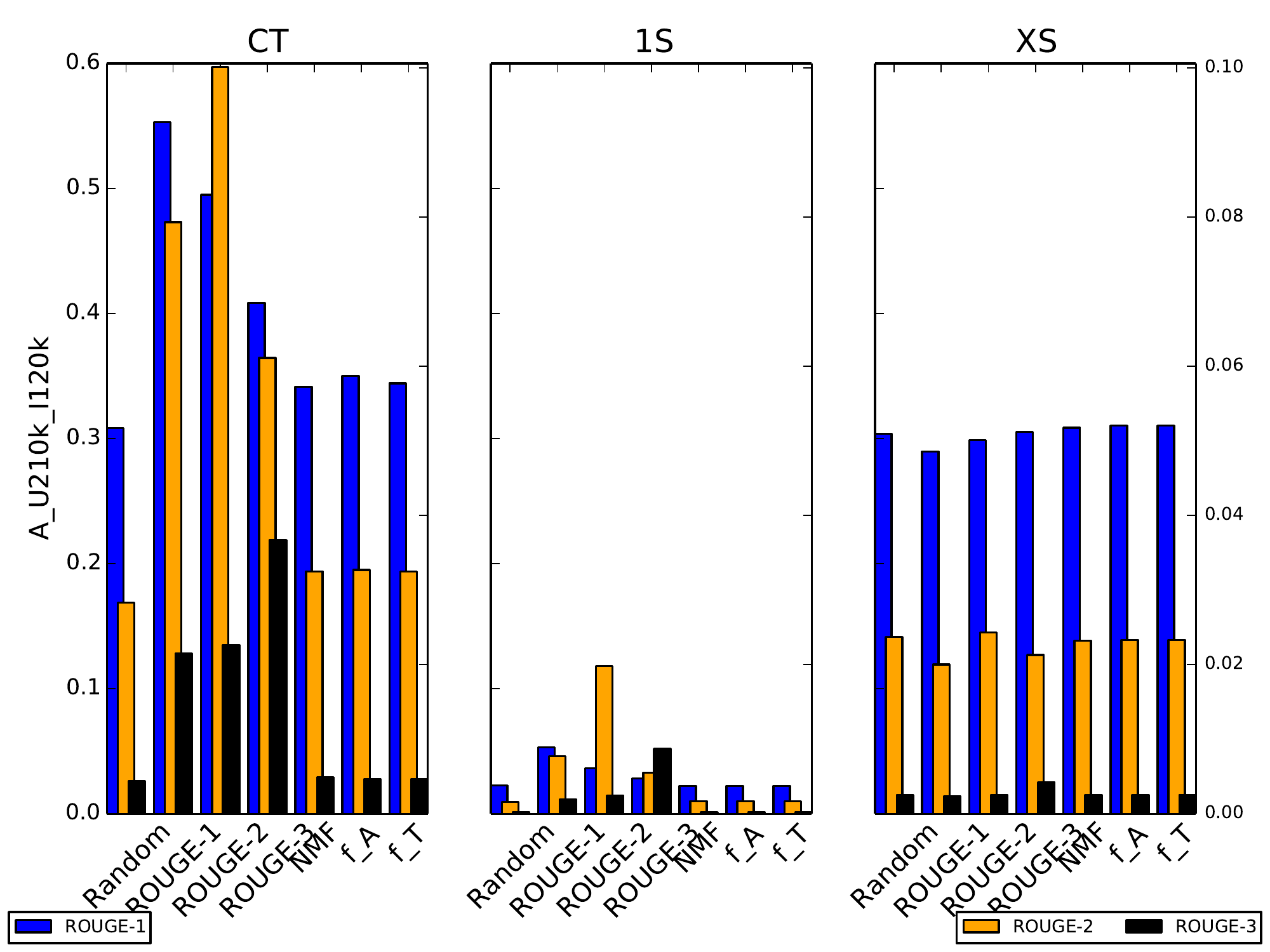}\caption{\label{fig:rouge amazon}Amazon experiments}\end{subfigure} 
\caption{\label{fig:rouge histograms} Histograms of the performances of the summarizer on the two biggest datasets. The scores of the ROUGE-1 metrics are represented in blue  while the scores of the ROUGE-2 and ROUGE-3 metrics are in yellow and black. 
7 models are compared: random, 3 oracles, NMF based model, $f_{A}$ and $f_{T}$ based models.
3 frameworks are investigated: CT (review extraction), 1S (One sentence extraction), XS (Multiple sentence extraction).
}
\end{figure*}

A first baseline is given by using a random scoring function $h$ (instead of the formulation given in~\eqref{eq:scoring model}). It provides a lower bound of the performance.
Three oracles are then used to provide an upper bound on the performance. They directly optimize the metrics ROUGE-1, ROUGE-2 and ROUGE-3 from the data on the test set.
 A matrix factorization baseline is also used. It is a special case of our model where no text information is used. This model computes a similar score for all the sentences of a given user and relative to an item. When one sentence only is selected, it is taken at random among the sentences of this user for the item. With greedy selection, the first sentence is chosen at random and then the cosine diversity term (algorithm \ref{algo:selsent}) allows a ranking of the next candidate sentences.
Our proposed method is evaluated with the two different user profile $\pi_u$ representation  (auto-encoder and raw text).
The performance of these seven models on the seven datasets with respect to the three metrics are aggregated in tables~\ref{tab:sumRB} and \ref{tab:sumAmazon} provided in the appendix.
A more synthetic version of the systems performance is provided through histograms in figure~\ref{fig:rouge histograms}. In this figure, one only shows the performance on the two biggest datasets.

An histogram corresponds to a text selection entity (whole review text, best single sentence, greedy sentence selection. Groups in the histograms (respectively row block of the tables) are composed of three cells corresponding respectively to the ROUGE-1, -2, -3 metrics. 
Not surprisingly, the results for the single sentence selection procedure (1S) are always worse than for the other two (CT: complete review  and XS: multiple sentences). This is simply because a sentence contains fewer words than a full review and it can hardly share more n-grams than the full text with the reference text. 
For the \emph{ratebeer.com} datasets, selecting a set of sentences clearly offers a better performance than selecting a whole review in all cases. 
Texts written to describe beers also describe the tasting experience. Was it in a bar or at home ? Was it a bottle or on tap ? Texts of the community share the same structure and vocabulary to describe both the tasting and the flavors of the beer. Most users write short and precise sentences. This is an appropriate context for our sentence scoring model, where the habits of users are caught by our recommender systems.
The performance slightly decreases when the size of the dataset is increased. 
As before, this is in accordance with the selection procedure of these datasets which focuses first on the most active users and commented items. 
For Amazon, the conclusion is not so clear and depending on the conditions, either whole reviews or selected sentences get the best score. It is linked to the higher variety in the community of users on the website: 
well structured sentences like those present in RateBeer are here mixed here with different levels of English and troll reviews.

The different models, overall, are following a clear hierarchy. 
First, stating the obvious, the random model has the worst performance.
Then, using a recommender system to select relevant sentences helps in terms of ROUGE-$n$ performance.
Using the text information brings most of the time only a small score improvement: the difference is however visible in tables~\ref{tab:sumRB} and~\ref{tab:sumAmazon}.
Overall our models only offer small improvements here with respect to random or NMF text selection. After analyzing this behavior, we believe that this is due to the shortness of the text reviews, to their relatively standardized form (arguments are very similar from one review to another), to the peaked vocabulary distribution of the reviews, and to the nature of ROUGE. The latter is a classical recall oriented summarization evaluation measure, but does not distinguishes well between text candidates in this context. This also shows that there is room for improvement on this aspect.

Concerning the oracle several conclusions can be drawn. For both single sentence and complete text selection, the gap between the ROUGE measures and the proposed selection method is important suggesting that there is still room for improvements here too. For the greedy sentence selection, the gap between the oracles and the hybrid recommender systems is moderate suggesting that the procedure is here fully efficient. However this conclusion should be moderated. It can be observed that whereas, ROUGE is effectively an upper bound for single sentence or whole review selection, this is no more the case for multiple sentences selection. Because of the complexity of selecting the best subset of sentences according to a loss criterion (which amounts at a combinatorial selection problem) we have been using a sub-optimal forward selection procedure: we first select the best ROUGE sentence, then the second best, etc. In this case the ROUGE procedure is no more optimal. 


Concerning the measures, the performance decreases rapidly when we move from ROUGE-1 to ROUGE-2 or ROUGE-3. 
Given the problem formulation and the context of short product reviews, ROUGE-2 and ROUGE-3 are clearly too constraining and the corresponding scores are not significant.

\subsection{Sentiment classification evaluation}
\label{sec:experiments:subsec:sentiment classification evaluation}

The performance of the different models, using the sentiment classification error as an evaluation metric, are presented in table~\ref{tab:ce all}.
Because they give very poor performance, the bias recommendation models ($\mu$ and $\mu_u$) are not presented here. The item bias $\mu_i$, second column, gives a baseline, which is improved by the matrix factorization $\gamma_u . \gamma_i$, third column. 
Our hybrid models $f_A$, fourth column, and $f_T$, fifth column, have lower classification errors than all the other recommender systems.
The first column, LL is the linear support vector machine (SVM) baseline.
It has been learnt on the training set texts, and the regularization hyperparameter has been selected using the validation set.
Our implementation relies on liblinear (LL) \cite{Fan:2008:LLL:1390681.1442794}.

Its performance is better than the recommender systems but it should be noted that it makes use of the actual text $d_{ui}$ of the review, whereas the recommender systems only use past information regarding user $u$ and item $i$. 
Note that even in this context, the recommender performance on RateBeer is very close to the SVM baseline.

It is then possible to combine the two models, according to the formulation proposed in section~\ref{sec:models:subsec:sentiment prediction model}.
The resulting hybrid approaches, denoted  LL + $f_A$ and LL + $f_T$, exploit both text based decision (SVM) and  user profile ($f_{A}$ and $f_{T}$).
This combined model shows good classification performance and overcomes the LL baseline in 4 out of 7 experiments in table \ref{tab:ce all}, while performing similarly to LL in the other 3 experiments.

\section{Overall gains}
\label{sec:overall gains}
\begin{figure*}[htpb!]
\centering
\begin{subfigure}[b]{0.3\textwidth}\includegraphics[scale=0.26]{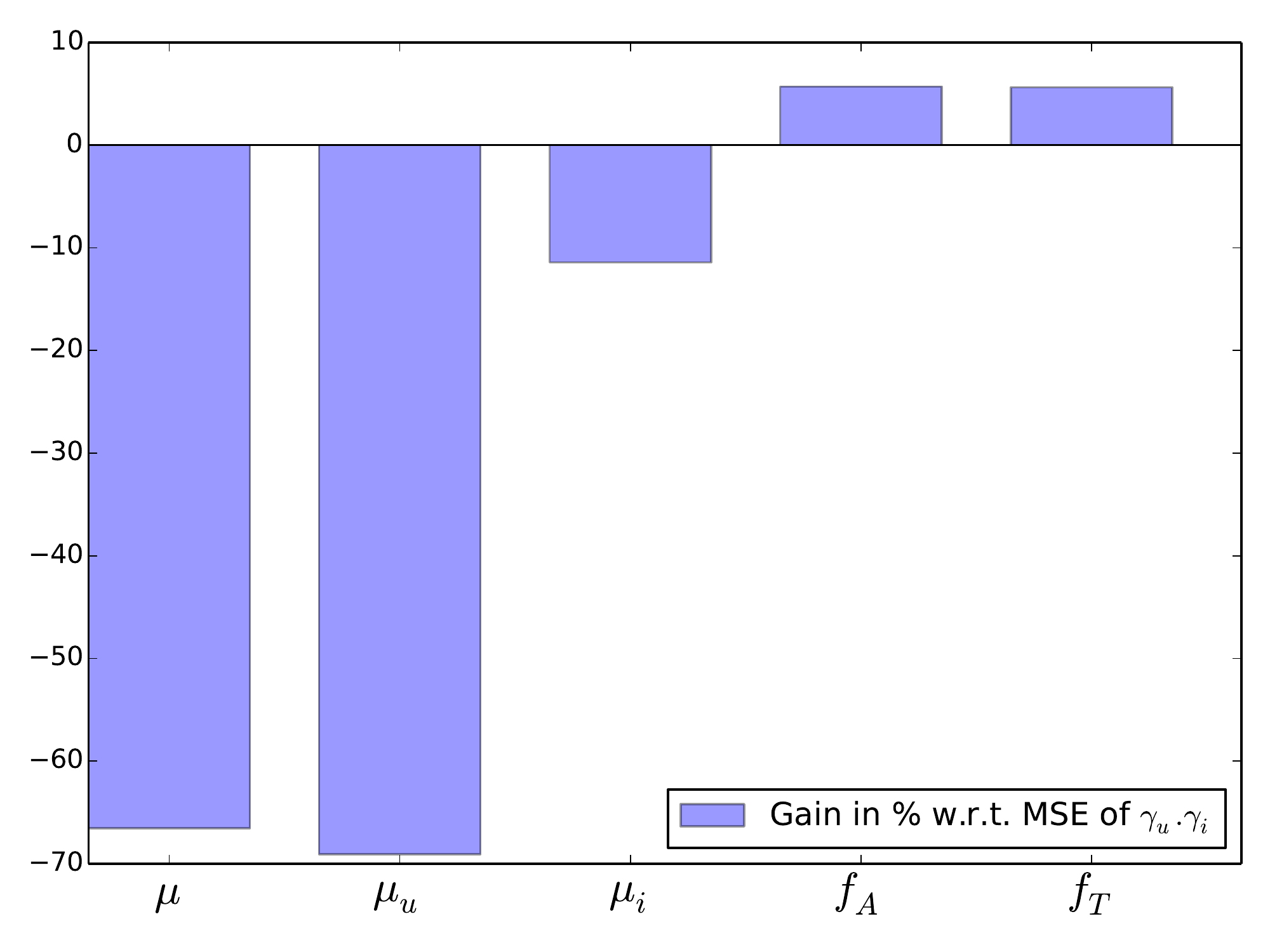}\caption{\label{fig:results recommender}Recommender systems. \\ Baseline=matrix factorization}\end{subfigure}
\begin{subfigure}[b]{0.3\textwidth}\includegraphics[scale=0.26]{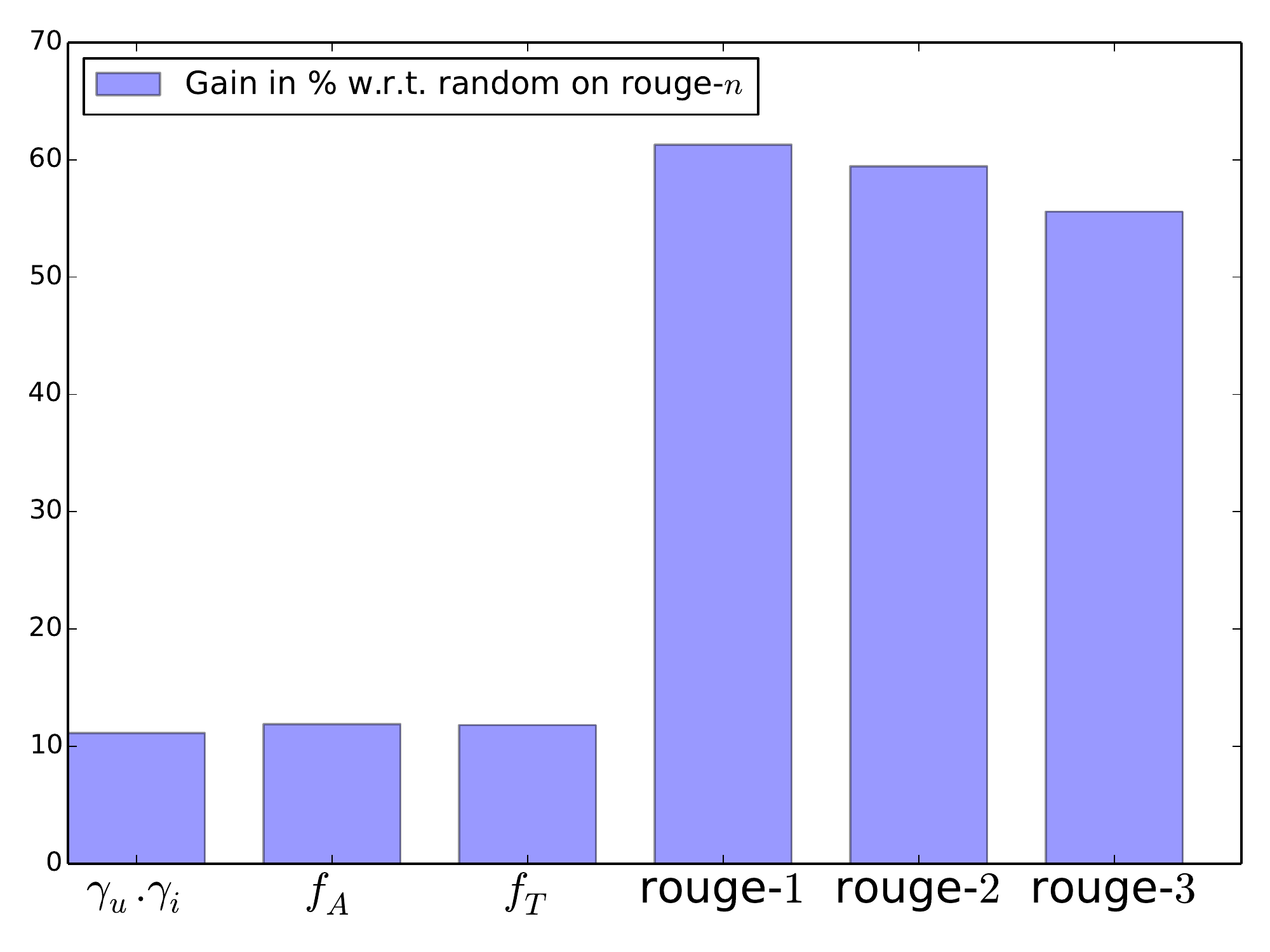}\caption{\label{fig:results summarizer}Summarizers. \\ Baseline=random selection procedure}\end{subfigure}
\begin{subfigure}[b]{0.3\textwidth}\includegraphics[scale=0.26]{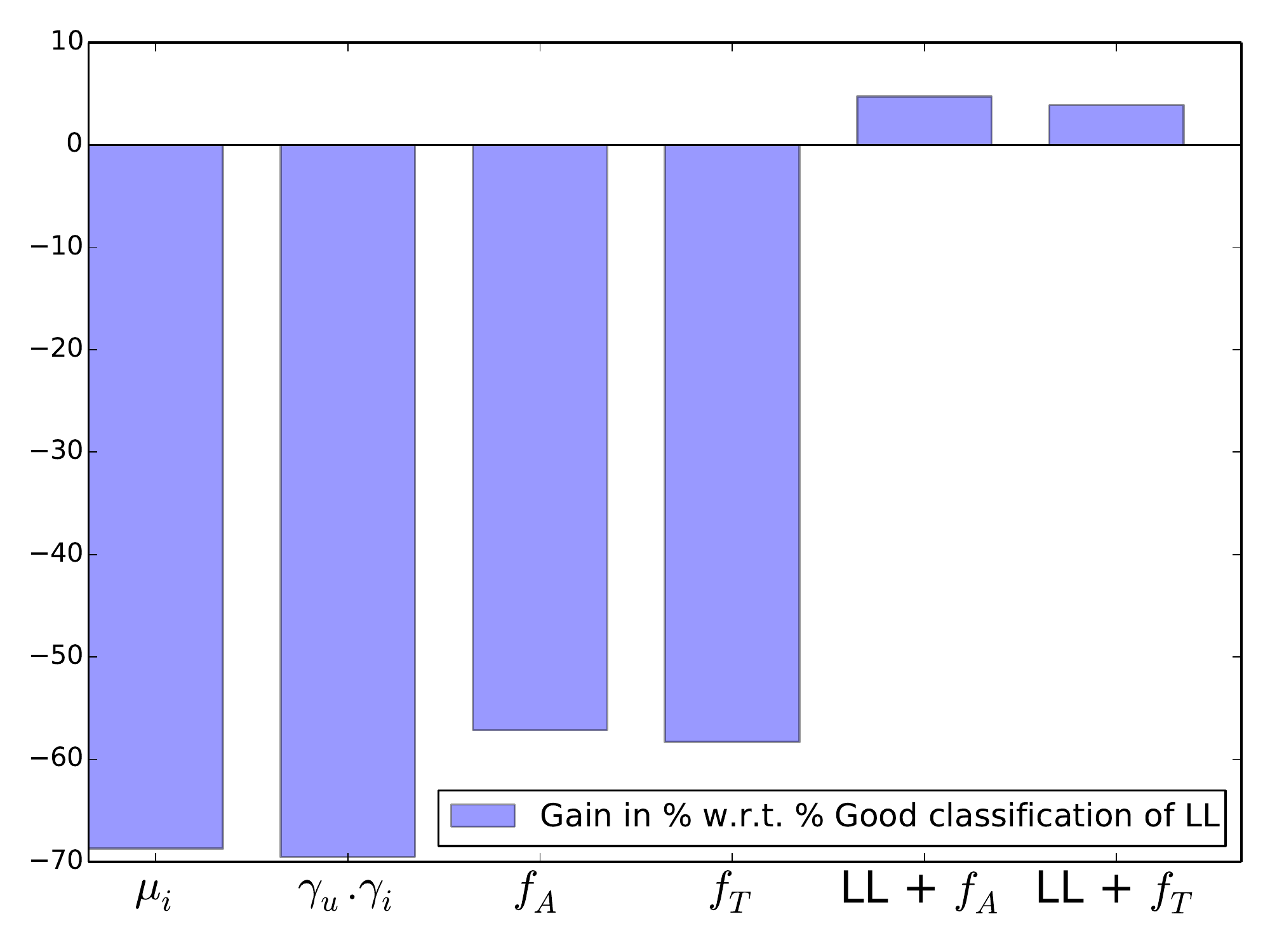}\caption{\label{fig:results opinion}Opinion classifiers. \\ Baseline=SVM}\end{subfigure}
\caption{\label{fig:results summary} Aggregated gains on the 3 tasks w.r.t. classic baselines: our hybrid recommender systems are better overall.}
\end{figure*}

In order to get a global vision of the overall gain provided by the proposed approach, we summarize here the results obtained on the different tasks. 
For each task, the gain with respect to the (task dependent) baseline is computed and averaged (per task) over all datasets. The metric depends on the task.
Results are presented  in figure~\ref{fig:results summary}.

For the mean squared error metric, presented in figure \ref{fig:results recommender}, the matrix factorization is used as baseline as it is the most common approach in the literature.
The user bias $\mu_u$ heavily fails to generalize on two datasets and has a negative gain: $-69.07\%$.
The item bias is closer to the baseline ($-11.43\%$).
Our hybrid models, which uses texts to refine user and item profiles bring a gain of $5.71\%$ for $f_A$, $5.63\%$ for $f_T$. This demonstrates the interest of including textual information in the recommender system.
Autoencoder and raw text approaches offer similar gains, the latter approach being overall faster.

For the text generation, we take the random model as baseline and results are presented in figure~\ref{fig:results summarizer}.
The gain is computed for the three investigated framework (CT: review selection, 1S: one sentence selection, XS: multiple sentence selection) and per measure (ROUGE-1, ROUGE-2 and ROUGE-3) and then averaged to one overall gain.
ROUGE-n oracles clearly outperform other models, which seems intuitive as they use ground truth for their predictions, with gains of $61.29\%$, $59.43\%$ and $55.59\%$.
The different recommender systems have very close behaviors with respective gains of $11.15\%$ (matrix factorization),  $11.89\%$ (auto-encoder), $11.83\%$ (raw text). Here textual information helps but does not clearly dominate ratings and provide only a small improvement. Remember that although performance improvement with respect to baselines is desirable, the main novelty of the approach here is to propose a personalized summary generation together with the usual rating prediction.


For the  opinion classifier, presented in figure~\ref{fig:results opinion}, the baseline consists in a linear SVM.
Basic recommender systems perform poorly with respect to the baseline (LL). 
Surprisingly, the item bias $\mu_i$ ($-68.71\%$) performs slightly better than matrix factorization $\gamma_u . \gamma_i$ ($-69.54\%$) in the context of sentiment classification (no neutral reviews and binary ratings).
Using textual information increases the performance. 
The autoencoder based model $f_A$ ($-57.17\%$) and raw text approach $f_T$ ($-58.31\%$) perform similarly.
As discussed in \ref{sec:experiments:subsec:sentiment classification evaluation}, the linear SVM uses the text of the current reviews when the recommender systems does not.
As a consequence, it is worth combining both predictions in order to exploit text and past profiles: the resulting models give respective gains of $4.72\%$ (autoencoder) and $3.89\%$ (raw text) w.r.t the SVM (LL).

\section{Related work}
\label{sec:related work}

Since the paper covers the topics of rating prediction, summarization and sentiment classification, we briefly present each of them in the following subsections and position ourselves in this context. We then discuss recent works combining or providing different sources of information.

\subsection{Recommender systems}
\label{sec:related work:subsec:recommender systems}

Three main families of recommendation algorithms have been developed \cite{burke2002hybrid}: content-based knowledge-based, and collaborative filtering\cite{Koren2009}. 
Given the focus of this work  on consumer reviews, we considered collaborative filtering.
These systems are intrinsically user-centered since they tackle the problem of selecting items to be recommended to a user given his past actions. For merchant websites the goal is to encourage users to buy new products and the problem is usually considered either as the prediction of a ranked list of relevant items for each user \cite{Breese1998,McLaughlin:2004:CFA:1008992.1009050} or as the completion of  missing ratings \cite{Resnick:1994:GOA:192844.192905,Koren2009}.
We have focused here on the latter approach for evaluation concerns: since we use data collected from third party sources (\emph{amazon.com} and \emph{ratebeer.com}) we would not able to evaluate dynamically the quality of a ranked list.
%

\subsection{Text summarization for consumer reviews}
\label{sec:related work:subsec:text summarization}
 
Early reference work \cite{Hu2006} on consumer reviews has focused on global summarization of user reviews for each item.
The motivation of this work was to extract the sentiments associated to a list of features from all the item review texts. The summarization took the form of a rating or of an appreciation of each feature.
Here, contrarily to this line of work, the focus is on personalized item summaries for a target  user.
Given the difficulty of producing a comprehensive  synthetic summary, we have turned this problem into a sentence or text selection process. 

 
Evaluation of summaries is challenging: how to assess the quality of a summary when the ground truth is subjective? 
In our context, the review texts are available and we used them as the ground truth. We have used classical ROUGE-$n$  summary evaluation measures \cite{lin2004rouge}.
For the evaluation, we take $n$ in the range 1 to 3.

\subsection{Sentiment classification}
\label{sec:related work:subsec:sentiment classification}

Different text latent representations have been proposed in this scope: 
 \cite{mei2007topic} proposed a generative model to  represent jointly topic and sentiments and recently, several works have considered  matrix factorization or neural network, in an attempts to develop robust sentiment recognition systems \cite{glorot2011domain}.
\cite{socher2012semantic} go further and propose to learn two types of representation: a vectorial model is learned for  word representation together with a latent transformation model, which allows the representation of negation and quantifiers associated to an expression.

We have investigated two kinds of representation for the texts:  bag of words and a latent representation through the use of autoencoders as in \cite{glorot2011domain}. \cite{McAuley2013a} also use a latent representation for representing reviews, although in a probabilistic setting instead in a deterministic one like we are doing here.



\subsection{Hybrid approaches}
\label{sec:related work:subsec:links between domains}

In the field of recommendation, a first hybrid model was proposed by \cite{Ganu2009}:  
it is based on hand labeling of review sentences (topic and polarity) to identify relevant characteristics of the items. Our approach does not need such hand labeling.
\cite{McAuley2013a} pushes further the exploitation of texts, by using a joint latent representation for ratings and textual content with the objective of improving the  rating accuracy. These two works are focused on rating prediction and do not consider delivering additional information to the user.
Very recently, \cite{zhang2014explicit} has considered adding an explanation component to a recommender system. 
For that, they propose to extract some keywords from the review texts, which are supposed to explain why a user likes or dislikes an item. This is probably the work whose spirit is closest to ours, but the components of their  system are only juxtaposed with no common variables of parameters and keyword generation is difficult to evaluate.

\cite{Hu2004,Hu2006} combined opinion mining and text summarization on product reviews with the goal of summarizing the qualities and defaults of the items.\cite{tan2012each} proposed a system for delivering personalized answers to user queries on specific products. They built the user profiles relying on topic modeling without any sentiment dimension. 
\cite{Agarwal2011} proposed a personalized news recommendation algorithm evaluated on the Yahoo portal using user feedback. 
This last reference is quite different from our setting and it does not investigate ratings or summarization issues.
Overall, we propose in this article to go beyond a generic summary of item characteristics by generating for each user a personalized summaries that is close to what they would have written about the item themselves.  

For a long time, sentiment classification has ignored the user dimension and has focused for example on the conception of \emph{"universal"} sentiment classifiers able to deal with a large variety of topics \cite{Blitzer07Biographies}.
Considering the user has become an  issue only very recently.
\cite{tan2011user} for example exploited explicit relations in social graphs for improving  opinion classifiers, but their work is only focused on this aspect.
\cite{McAuley2013b} proposed to  distinguish different rating behaviors and show that modeling the review authors in a scale ranging from connoisseur to expert offers a significant gain for an opinion classification task. Again they focused on sentiment.

In our work, we have experimented the benefits of considering the text of user reviews in recommender system for their performance as sentiment classifier. We have additionally proposed, as a secondary contribution, an original model mixing recommender systems and linear classification. In another work, they integrate the textual data written by each user in the recommendation process \cite{McAuley2013a}. In this sense, their work is close to our rating prediction model, although their formalism is different. However, they do not consider offering the user additional information (e.g. text reviews) besides ratings.



\renewcommand\thetable{A\arabic{table}}
\begin{table*}[htb!]
\centering
\scriptsize
\begin{tabular}{|c|c||c|c|c||c|c|c||c|c|c|}
	\hline
	&Dataset                      & \multicolumn{3}{c||}{RB\_U50\_I200}&  \multicolumn{3}{c||}{RB\_U500\_I2k} &  \multicolumn{3}{c|}{RB\_U5k\_I20k} \\
	&Performance measure                       & R-1  & R-2  & R-3   & R-1  & R-2  & R-3  & R-1  & R-2  & R-3   \\
	\hline
	\multirow{7}{*}{\rotatebox{90}{Complete Text}} &Random review                             & 0.2339 & 0.0160 & 0.0018  & 0.2321 & 0.0150 & 0.0014  & 0.2190 & 0.0143 & 0.0016  \\
	&Best ROUGE-1 review                       & 0.4903 & 0.0843 & 0.0489  & 0.5463 & 0.0976 & 0.0571  & 0.5334 & 0.0900 & 0.0501\\
	&Best ROUGE-2 review                       & 0.3693 & 0.1307 & 0.0512  & 0.3995 & 0.1614 & 0.0614  &0.3849 & 0.1567 & 0.0548 \\
	&Best ROUGE-3 review                       & 0.3106 & 0.0730 & 0.0722  &0.3263 & 0.0892 & 0.1022  &0.3108 & 0.0819 & 0.0968\\
	&$\gamma_u . \gamma_i $ (best review)      & 0.2499 & 0.0178 & 0.0013  & 0.2317 & 0.0159 & 0.0015 & 0.2191 & 0.0147 & 0.0017\\
	&$f_A$ (best review)                       & 0.2543 & 0.0183 & 0.0013  &  0.2334 & 0.0160 & 0.0015  &0.2204 & 0.0148 & 0.0016 \\
	&$f_T$ (best review)                       & 0.2543 & 0.0182 & 0.0013  & 0.2334 & 0.0160 & 0.0015  & 0.2206 & 0.0148 & 0.0017\\
	\hline 
	\multirow{7}{*}{\rotatebox{90}{Single  sentence}}&Random sentence                           & 0.0524 & 0.0026 & 0.0002  & 0.0429 & 0.0026 & 0.0002  &0.0440 & 0.0026 & 0.0003\\
	&Best ROUGE-1 sentence                     & 0.1486 & 0.0221 & 0.0071  &0.1490 & 0.0256 & 0.0079  &   0.1569 & 0.0271 & 0.0087 \\
	&Best ROUGE-2 sentence                     & 0.0971 & 0.0587 & 0.0080  & 0.1012 & 0.0704 & 0.0102  & 0.1061 & 0.0740 & 0.0118\\
	&Best ROUGE-3 sentence                     & 0.0724 & 0.0151 & 0.0215  & 0.0780 & 0.0228 & 0.0429  &0.0805 & 0.0241 & 0.0441\\
	&$\gamma_u . \gamma_i $ (best sentence)    & 0.0557 & 0.0045 & 0.0001  & 0.0455 & 0.0034 & 0.0003  & 0.0471 & 0.0036 & 0.0004 \\
	&$f_A$ (best sentence)                     & 0.0556 & 0.0043 & 0.0002  & 0.0456 & 0.0034 & 0.0003  & 0.0472 & 0.0036 & 0.0004\\
	&$f_T$ (best sentence)                     & 0.0557 & 0.0043 & 0.0002  & 0.0456 & 0.0034 & 0.0003  &  0.0471 & 0.0035 & 0.0004\\
	\hline
	\multirow{7}{*}{\rotatebox{90}{Set of sentences}}&Random greedy sel.                  & 0.2785 & 0.0163 & 0.0005  &0.2508 & 0.0115 & 0.0008  & 0.2437 & 0.0121 & 0.0011 \\
	&ROUGE-1 greedy sel.                 & 0.5088 & 0.0576 & 0.0092  &0.5088 & 0.0576 & 0.0092  & 0.1470 & 0.0075 & 0.0013 \\
	&ROUGE-2 greedy sel.                 & 0.3126 & 0.0632 & 0.0062  &0.3126 & 0.0632 & 0.0062  &0.2549 & 0.0138 & 0.0019   \\
	&ROUGE-3 greedy sel.                 & 0.3972 & 0.0299 & 0.0150  & 0.3972 & 0.0299 & 0.0150  &0.3210 & 0.0154 & 0.0031\\
	&$\gamma_u . \gamma_i $ (sentence, greedy) & 0.4251 & 0.0313 & 0.0053  &0.3450 & 0.0171 & 0.0012  &  0.3428 & 0.0176 & 0.0015 \\
	&$f_A$ (sentence, greedy)                  & 0.4248 & 0.0316 & 0.0041  & 0.3484 & 0.0173 & 0.0012  &0.3426 & 0.0177 & 0.0015 \\
	&$f_T$ (sentence, greedy)                  & 0.4247 & 0.0314 & 0.0041  & 0.3482 & 0.0173 & 0.0012  & 0.3446 & 0.0177 & 0.0014 \\
	\hline
\end{tabular}
\caption{ROUGE-n evaluation on RateBeer subsets. Top columns are different datasets (see text); R-$n$ is ROUGE-$n$ measure. Row blocks represent generating procedures (CT, 1S, XS). Each row corresponds to a text prediction model.}
\label{tab:sumRB}
\end{table*}
\setlength{\tabcolsep}{.5em}
\begin{table*}[tb!]
\centering
\scriptsize
\begin{tabular}{|c|c||c|c|c||c|c|c||c|c|c||c|c|c|}
	\hline
	&Dataset                      & \multicolumn{3}{c||}{A\_U200\_I100} & \multicolumn{3}{c||}{A\_U2k\_I1k} & \multicolumn{3}{c||}{A\_U20k\_I10k}  &\multicolumn{3}{c|}{A\_U200k\_I100k} \\
	&Performance measure                       & R-1  & R-2  & R-3   & R-1  & R-2  & R-3 & R-1  & R-2  & R-3 & R-1  & R-2  & R-3   \\
	\hline
	\multirow{7}{*}{\rotatebox{90}{Complete Text}}&Random review                             &  0.3786 & 0.0393& 0.0038  &  0.3350 & 0.0337 & 0.0047&  0.3213 & 0.0304 & 0.0039 & 0.3085 & 0.0283 & 0.0044 \\
	&Best ROUGE-1 review                       &0.6059 & 0.0957 & 0.0245  &  0.6030 & 0.0918 & 0.0204& 0.5945 & 0.0882 & 0.0205 & 0.5530 & 0.0793 & 0.0215\\
	&Best ROUGE-2 review                       & 0.5642 & 0.1076 & 0.0260&  0.5511 & 0.1116 & 0.0213 & 0.5371 & 0.1104 & 0.0218& 0.4950 & 0.1001 & 0.0226 \\
	&Best ROUGE-3 review                       & 0.4897 & 0.0815 & 0.0339&  0.4510 & 0.0691 & 0.0368  & 0.4345 & 0.0661 & 0.0385& 0.4085 & 0.0611 & 0.0367\\
	&$\gamma_u . \gamma_i $ (best review)      &  0.3944 & 0.0467 & 0.0041  &  0.3525 & 0.0358 & 0.0050& 0.3379 & 0.0325 & 0.0042 & 0.3414 & 0.0325 & 0.0049\\
	&$f_A$ (best review)                       &  0.4118 & 0.0468 & 0.0046  & 0.3546 & 0.0365 & 0.0051 & 0.3385 & 0.0326 & 0.0042 & 0.3501 & 0.0327 & 0.0046 \\
	&$f_T$ (best review)                       &  0.4124 & 0.0468 & 0.0045  & 0.3552 & 0.0366 & 0.0051 & 0.3385 & 0.0326 & 0.0042& 0.3441 & 0.0325 & 0.0046  \\
	\hline 
	\multirow{7}{*}{\rotatebox{90}{Single  sentence}}&Random sentence                           &  0.0226 & 0.0023 & 0.0006  &  0.0180 & 0.0012 & 0.0001 & 0.0199 & 0.0014 & 0.0001& 0.0226 & 0.0016 & 0.0002 \\
	&Best ROUGE-1 sentence                     &  0.0435 & 0.0047 & 0.0007 &  0.0437 & 0.0063 & 0.0016& 0.0496 & 0.0077 & 0.0020 & 0.0531 & 0.0077 & 0.0019\\
	&Best ROUGE-2 sentence                     &  0.0304 & 0.0170 & 0.0018 & 0.0303 & 0.0181 & 0.0024& 0.0341 & 0.0205 & 0.0027 & 0.0363 & 0.0198 & 0.0024\\
	&Best ROUGE-3 sentence                     &  0.0210 & 0.0035 & 0.0041  & 0.0241 & 0.0054 & 0.0086 & 0.0265 & 0.0061 & 0.0100 & 0.0283 & 0.0055 & 0.0087\\
	&$\gamma_u . \gamma_i $ (best sentence)    &  0.0199 & 0.0013 & 0.0004 & 0.0181 & 0.0016 & 0.0001 & 0.0195 & 0.0015 & 0.0002  & 0.0222 & 0.0017 & 0.0002 \\
	&$f_A$ (best sentence)                     &  0.0191 & 0.0016 & 0.0005 & 0.0181 & 0.0016 & 0.0001& 0.0196 & 0.0015 & 0.0002& 0.0222 & 0.0017 & 0.0002  \\
	&$f_T$ (best sentence)                     &  0.0191 & 0.0016 & 0.0005  & 0.0182 & 0.0016 & 0.0001& 0.0196 & 0.0015 & 0.0002 & 0.0222 & 0.0017 & 0.0002\\
	\hline
	\multirow{7}{*}{\rotatebox{90}{Set of sentences}}&Random greedy sel.                  & 0.3518 & 0.0325 & 0.0025 &  0.3753 & 0.0323 & 0.0035& 0.3613 & 0.0298 & 0.0030 & 0.3038 & 0.0237 & 0.0025\\
	&ROUGE-1 greedy sel.                 &  0.3747 & 0.0338 & 0.0022  & 0.3625 & 0.0253 & 0.0024& 0.3440 & 0.0234 & 0.0024 & 0.2896 & 0.0200 & 0.0023\\
	&ROUGE-2 greedy sel.                 &  0.3615 & 0.0430 & 0.0029 &  0.3650 & 0.0259 & 0.0024& 0.3544 & 0.0259 & 0.0026  & 0.2988 & 0.0243 & 0.0025\\
	&ROUGE-3 greedy sel.                 & 0.3571 & 0.0341 & 0.0048 &  0.3730 & 0.0272 & 0.0043& 0.3665 & 0.0260 & 0.0043& 0.3054 & 0.0213 & 0.0042\\
	&$\gamma_u . \gamma_i $ (sentence, greedy) &  0.3615 & 0.0329 & 0.0034 & 0.3785 & 0.0309 & 0.0032 & 0.3710 & 0.0296 & 0.0030& 0.3087 & 0.0232 & 0.0025\\
	&$f_A$ (sentence, greedy)                  &  0.3589 & 0.0331 & 0.0029 &  0.3794 & 0.0313 & 0.0034 & 0.3726 & 0.0298 & 0.0030& 0.3103 & 0.0233 & 0.0025 \\
	&$f_T$ (sentence, greedy)                  &  0.3586 & 0.0332 & 0.0029 &  0.3792 & 0.0312 & 0.0033 & 0.3726 & 0.0298 & 0.0030 & 0.3103 & 0.0233 & 0.0025\\
	\hline
\end{tabular}
\caption{ROUGE-n evaluation on Amazon subsets. Top columns are different datasets (see text); R-$n$ is ROUGE-$n$ measure. Row blocks represent generating procedures (CT, 1S, XS). Each row corresponds to a text prediction model.}
\label{tab:sumAmazon}
\end{table*}

\section{Conclusion}
\label{sec:conclusion}
This article proposes an extended framework to the recommendation task. The general goal is to enrich classical recommender systems with several dimensions. As an example we show how to generate personalized reviews for each recommendation using extracted summaries. This is our main contribution. We also show how rating and text could be used to produce efficient personalized sentiment classifiers for each recommendation. Depending on the application, other additional information could be brought to the user.
Besides producing additional information for the user, the different information sources can take benefit one from the other. We thus show how to effectively make use of text review and rating informations for building improved rating predictors and review summaries. As already mentioned, the sentiment classifiers also benefits from the two information sources. This part of the work demonstrates that multiple information sources could be useful for improving recommendation systems. This is particularly interesting since several sources are effectively available now at many online sites.
Several new applications could be developed along the lines suggested here. From a modeling point of view, more sophisticated approaches can be developed. We are currently working on a multitask framework where the representations used in the different components are more closely correlated than in the present model.

\bibliography{wsdm2015_paper}{}
\bibliographystyle{plain}
\begin{appendix}
	\section{Result tables for ROUGE-n metrics} 
	\label{app:result tables}
	This appendix gathers the result tables of the ROUGE-n metrics on all datasets, for all prediction procedures and all models as described in section \ref{sec:experiments:subsec:text generation evaluation}. 
	Performance on all datasets extracted from  \emph{ratebeer.com} and \emph{amazon.com}  are in tables \ref{tab:sumRB} and \ref{tab:sumAmazon} respectively.  
	Full name are used instead of abbreviation in the first column of the table to name each generation procedure. The abbreviations were CT, 1S and XS for respectively Complete Text, Single Sentence and Set of sentences in text and in the histograms \ref{fig:rouge histograms}.
	Description of the size of the different datasets is available in \ref{tab:datasets information}, the datasets grow in number of reviews (and users and items) from left to right. Greedy refers to our selection algorithm \ref{algo:selsent}, presented in section \ref{sec:models:subsec:text generation model}.
\end{appendix}

\end{document}